\documentclass[%
showpacs,preprintnumbers,
 amsmath,amssymb,
 aps,
prb,
]{revtex4}

\usepackage{graphicx}
\usepackage{dcolumn}
\usepackage{bm}

\usepackage{amsmath}    
\usepackage{graphicx}   
\usepackage{verbatim}   
\usepackage{color}      
\usepackage{subfigure}  
\usepackage{hyperref}   
\usepackage{mathrsfs}

\begin{document}


\title{A new class of scalable parallel pseudorandom number generators based on Pohlig--Hellman exponentiation ciphers}

\author{Paul D. Beale}
\email{paul.beale@colorado.edu}
\affiliation{%
University of Colorado Boulder 
}%

\date{\today}

\begin{abstract}
Parallel supercomputer-based Monte Carlo applications depend on pseudorandom number generators  that produce independent pseudorandom streams across many separate processes. 
We propose a new scalable class of parallel pseudorandom number generators  based on Pohlig--Hellman  exponentiation ciphers. The method generates uniformly distributed floating point pseudorandom streams by encrypting simple sequences of integer \textit{messages} into \textit{ciphertexts} by exponentiation modulo prime numbers.  The advantages of the method are:  the method is trivially parallelizable by parameterization with each pseudorandom number generator derived from an independent prime modulus, the method  is fully scalable on massively parallel computing clusters due to the large number of primes available for each implementation, the seeding and initialization of the independent  streams is simple, the method requires only a few integer multiply--mod operations per pseudorandom number, the state of each instance is defined by only a few integer values, the period of each instance is different, 
and the method passes a battery of intrastream and interstream correlation tests using up to $10^{13}$ pseudorandom numbers per test. The 32-bit implementation we propose has millions of possible instances, all with periods greater than $10^{18}$.  A 64-bit implementation depends on 128-bit arithmetic, but would have more than $10^{15}$ possible instances and periods greater than $10^{37}$. 
\end{abstract}

\maketitle


\section{\label{sec:Introduction}Introduction}

We propose a new class of scalable parallel pseudorandom number generators for use in large-scale Monte Carlo and other computer applications that require large numbers of independent streams of pseudorandom numbers. The method we propose is
based on Pohlig--Hellman exponentiation ciphers.\cite{PohligHellman1978, 
Schneier1994} 
 The method creates a pseudorandom stream by encrypting a simple sequence of short integer plaintext \textit{messages} $m_k$ into \textit{ciphertexts} $c_k$ using the transformation 
\begin{align}\label{eqn:encrypt2}
c_k&= m_k^e \thinspace\textrm{mod}\thinspace  n ,
\end{align}
where each generator instance is based on an independent prime modulus $n$, and an exponent $e$ that is co-prime to $n-1$.
Here and throughout, $x=y \thinspace\textrm{mod}\thinspace z$ means $x$ is the remainder of $y$ upon division by $z$,  with $0 \leq x < z$.
The method is based upon elementary number theory.\cite{Koshy2002, Silverman2006, Koblitz1987}  For every prime number $n$, the set of integers $\mathbb{Z}/n\mathbb{Z} = [0 \thinspace . \thinspace . \thinspace n-1]$ forms a finite field that is closed under addition and multiplication modulo $n$. Also, the set of nonzero elements $(\mathbb{Z}/n\mathbb{Z})^* = [1 \thinspace . \thinspace . \thinspace n-1]$ forms a group that is closed under multiplication, and for every element $a$ in $(\mathbb{Z}/n\mathbb{Z})^*$ there exists a unique inverse $a^{-1}$ such that $a a^{-1}\thinspace\textrm{mod}\thinspace  n = 1$ . The pseudorandom number generator algorithm described here cycles through a sequence of messages $m_k$ with $k=0,1,2,\ldots$ selected from $\mathbb{Z}/n\mathbb{Z}$ using some simple pattern that uniformly samples the set. The exponentiation step equation \eqref{eqn:encrypt2} then gives a 
sequence of ciphertexts $c_k$ that is uniformly distributed on $\mathbb{Z}/n\mathbb{Z}$. Uniformly distributed double precision floating point values $R_k$ on the open real interval $(0 \thinspace . \thinspace . \thinspace 1)$ are formed with a floating point division: $R_k=(c_k+1) / (n+1)$. 

Most pseudorandom number generators generate the next pseudorandom integer from either the previous pseudorandom integer in the sequence, or by using two or more pseudorandom integers from earlier in the sequence. This method is different in that the pseudorandom sequence arises from the encryption of a  sequence of integer messages. In this way, it is similar to cryptographically secure pseudorandom number generators\cite{Schneier1994, FergusonSchneierKohno2010} and floating point pseudorandom number generators\cite{NumericalRecipes1992} based on block ciphers. The quality of the pseudorandom sequence from our method is based on modular exponentiation being a good one-way cryptographic function, meaning that it is computationally difficult to ascertain the message from the ciphertext without knowing both the modulus and exponent.\cite{Schneier1994, FergusonSchneierKohno2010, Koblitz1987, BlumMichali, PatelSundaram} The algorithm we propose here produces 
excellent, long-period pseudorandom sequences even for 32-bit prime moduli. The period of the 32-bit implementation we propose is greater than $10^{18}$, and has  millions of possible instances. Sixty-four bit implementations have periods in excess of $10^{37}$, with more than $10^{15}$ possible instances.

Two qualitatively different schemes have been used to create scalable systems of pseudorandom number generators for use on massively parallel supercomputers: stream splitting and parameter splitting.\cite{BaukeMertens2007} Stream splitting is based on a single pseudorandom number generator with an extremely long period, with parallelization  accomplished by subdividing the full period into independent non-overlapping subsequences. This method requires careful seeding to ensure that no two subsequences will overlap for any feasible set of processes.  
Parameter splitting uses a single algorithm, but produces independent pseudorandom sequences by assigning different parameters to each process.
The SPRNG web site\cite{MascagniSrinivasan2000,sprng} gives examples of parallel generators of both classes. For example, linear congruential generators of the form $s_k= ( a s_{k-1} + b ) \thinspace\textrm{mod}\thinspace p$, with $p$ a power of two, can be parallelized by parameter splitting.\cite{lecuyer1999, MascagniSrinivasan2000,sprng, mascagni1998} The parameters are $a$, a well-tested multiplier that ensures a maximal period and $b$, an odd number less than $p$. Each process uses the same values of $p$ and $a$, but the value of $b$ is chosen to be different for each process.\cite{poweroftwocongruentialweaknesses} Another widely used class of parallel generators are based on the lagged Fibonacci method.\cite{mascagni1995a, mascagni1995b, MascagniSrinivasan2000, sprng, MersenneTwister}  
They are usually of the form $s_k = (s_{k-q} \otimes s_{k-r}) \thinspace\textrm{mod}\thinspace p$, where $p$ is a power of two, $\otimes$ is one of the operations addition, exclusive or,  subtraction, or multiplication, and $q<r$ are integer parameters chosen based on primitive polynomials modulo 2.\cite{Schneier1994, Koblitz1987, knuth} Lagged Fibonacci generators have periods of the order of $2^r$, with $r$ typically of the order of several hundred to several thousand. The state of the generator is defined by a table of $r$ integers or $r$ bits, which represent the most recent pseudorandom values in the sequence. Parallel implementation of these algorithms can be accomplished by either stream splitting or parameter splitting.
\cite{MascagniSrinivasan2000, sprng}

The parallel pseudorandom number generator class we propose here is based on parameter splitting. In our algorithm, every independent process is assigned a different prime modulus $n$, which produces an independent pseudorandom sequence. The sequence of pseudorandom numbers produced using equation \eqref{eqn:encrypt2}  are effectively uncorrelated within a single stream and between streams, and the period of every stream is different. The number of independent streams is limited only by the number of prime numbers in the range defined by the implementation. For example, in a 32-bit implementation the prime moduli can be chosen from the set of 98,182,656 primes in the range $[2^{31}  . \thinspace . \thinspace\thinspace 2^{32}]$.\cite{http://oeis.org/A036378}

To demonstrate the relation to encryption and explain some of the useful properties of the method, the message $m$ can be decrypted from the ciphertext $c$ using the decryption exponent $d=e^{-1}\thinspace \textrm{mod}\thinspace (n-1)$, i.e.
 $ d e = 1+ q (n-1) $ for some $q$.\cite{PohligHellman1978, Schneier1994, Koshy2002, Silverman2006, Koblitz1987}
The decryption exponent $d$ exists and is unique if $e$ and $n-1$ are co-prime, i.e if $\gcd(e,n-1)=1$. The decryption exponent can be determined quickly using the extended Euclidean algorithm.\cite{Schneier1994, Koshy2002, Silverman2006, Koblitz1987}  
The decryption is based on Fermat's little theorem:\cite{Koshy2002, Silverman2006, Koblitz1987} 
for any prime n and for all $m$ in $(\mathbb{Z}/n\mathbb{Z})^*$, $m^{n-1}\thinspace\textrm{mod}\thinspace n= 1$. Therefore 
\begin{align}\label{eqn:decrypt}
c^d \thinspace \textrm{mod} \thinspace n = m^{d e} \thinspace \textrm{mod} \thinspace n =
m^{1+q(n-1)} \thinspace \textrm{mod} \thinspace n =
(m (m^{n-1})^q) \thinspace \textrm{mod} \thinspace n =
m \thinspace \textrm{mod} \thinspace n = m .
\end{align}
For any allowed exponent $e$, equation \eqref{eqn:encrypt2} maps $\mathbb{Z}/n\mathbb{Z}$ onto a permutation of the same set. Therefore, 
any message sequence that uniformly samples $\mathbb{Z}/n\mathbb{Z}$ will produce ciphertexts that uniformly sample $\mathbb{Z}/n\mathbb{Z}$. The Pohlig--Hellman algorithm is closely related to the more widely used RSA public key encryption method\cite{Schneier1994, Koshy2002, Silverman2006, Koblitz1987, FergusonSchneierKohno2010, rsa1979} which uses moduli that are products of two large primes. Using a composite modulus $n=p q$ creates a serous weakness for this application since every message $m$ that is a multiple of $p$ or $q$ gives a ciphertext $c$ that is also a multiple of $p$ or $q$, respectively.
 
In the Pohlig--Hellman cryptographic scheme, the key consisting of the prime $n$ and exponent $e$ must be known only to the sender and receiver. Otherwise, an eavesdropper can easily determine the decryption exponent. Cryptography theory suggests using safe primes for the moduli, i.e.~primes $n$ for which $(n-1)/2$ is also prime.\cite{PohligHellman1978, PatelSundaram, Schneier1994, SophieGermain} This helps ensure eavesdroppers will find it computationally difficult to take the discrete logarithm 
to recover the message from the ciphertext.\cite{PohligHellman1978, Koshy2002, Silverman2006, Koblitz1987} While our empirical tests do not give noticeably better statistics for safe primes than for primes in general, the number of safe primes does not seriously limit the scaling with 32-bit moduli since there are 3,060,794 safe primes in the range $[2^{31}  . \thinspace . \thinspace \thinspace 2^{32}]$.\cite{http://oeis.org/A211395} 

The algorithm presented here is similar to a block cipher operating in the counter mode which outputs a sequence of pseudorandom numbers that are ciphertexts resulting from a simple sequence of plaintexts.\cite{Schneier1994, FergusonSchneierKohno2010} 
The Pohlig--Hellman method can be used to generate cryptographically secure pseudorandom sequences, but cryptographic applications require primes with hundreds of digits.\cite{PohligHellman1978, Schneier1994, Koshy2002, Silverman2006, Koblitz1987, FergusonSchneierKohno2010, BlumMichali, PatelSundaram}  
Our application, instead, uses Pohlig--Hellman encryption of 32-bit 
messages to produce pseudorandom sequences suitable for use in Monte Carlo simulations and other applications. 
The exponentiations can be accomplished using the method of repeated squaring,\cite{Schneier1994, Koshy2002, Silverman2006, Koblitz1987} so the number of multiply--mod operations needed to calculate $m^e\thinspace\textrm{mod}\thinspace n$ is less than $2\log_2 e$. As we will show below, the algorithm can be implemented using small exponents that require very few multiply--mod operations per pseudorandom number.\cite{pohlig-hellansmallexponents}
If the modulus $n$ is chosen to be a prime number less than $2^{32}$, each multiply--mod operation can be performed in hardware on a 64-bit processor with one 64-bit multiply, and one 64-bit mod. Our algorithm leads to a new scalable class of fast and effective pseudorandom generators based on parameter splitting. 
There are several advantages to this pseudorandom number generation method.

\begin{itemize}

\item The algorithm is based on elementary number theory and cryptography.

\item The method is trivially parallelizable by parameterization, with each instance derived from an independent prime modulus. 

\item Pseudorandom sequences that result from different prime moduli are independent, and  have different  periods.

\item The method is fast since it requires only a few integer multiply--mod operations per pseudorandom number.

\item The method is fully scalable on massively parallel computing clusters due to the millions of available 32-bit primes. 

\item As we will show below, the period of the generator can be greatly extended by using message skips derived from a prime number linear congruential pseudorandom number generator.

\item The seeding and initialization of the independent  streams is simple, and it is possible to initialize each process without needing information about the states of any of the other processes.
 The state of each generator is defined by only seven integers: $m$, $e$, $n$, and $c$, and integer pseudorandom skip parameters $s$, $p$ and $a$ defined below.  All of these values can be stored in local memory, and require no global storage. The values $n$, $e$, $p$ and $a$ are fixed in each instance. Only the three values $m$, $s$, and $c$ change during calls to the generator.

\item The algorithm allows one to quickly jump far forward or backward in the pseudorandom sequence. 

\item The method does not require combining generators\cite{lecuyer1988} or shuffling\cite{BayesDurham} to remove correlations.

\item The algorithm is simple enough to allow the generator to be implemented as an in-line function for efficiency. 

\item The algorithm can be implemented in parallel on vector computers and GPUs.

\item The method passes a battery of intrastream and interstream correlation tests using up to $10^{13}$ pseudorandom numbers per test. 

\end{itemize}

\section{Message skipping algorithms}

The selection of messages to be encrypted for a given modulus $n$ can be accomplished in many different ways, but they can all be expressed in terms of an integer skip sequence with the skips $s_k$ 
chosen from $\mathbb{Z}/n\mathbb{Z}$:
\begin{subequations}\label{basealgorithm}
\begin{align}
m_k&= (m_{k-1}+s_k) \thinspace\textrm{mod}\thinspace n,\\
c_k &= m_k^e \thinspace\textrm{mod}\thinspace n .
\end{align}
\end{subequations}
Note that if the skip values $s_k$ are chosen uniformly and \textit{randomly} (not pseudorandomly) from $\mathbb{Z}/n\mathbb{Z}$, then the sequence of messages $m_k$ forms a uniform random sequence on the same set. Message $m_k$ is, in effect, a one-time pad encryption of message $m_{k-1}$.\cite{Schneier1994}  Since the set of ciphertexts is one-to-one with the set of messages, the sequence of ciphertexts $c_k$ also forms a  uniformly distributed random sequence. 

In encryption, it is essential to avoid \textit{cribs}, i.e. messages that result in easily decoded ciphertexts. For example, the messages $m=0,1,n-1$ are cribs for all allowed exponents $e$ since $m^e \thinspace\textrm{mod}\thinspace n =0,1,n-1$, respectively. 
Calculation efficiency makes it desirable to use small exponents to reduce the number of multiply--mod operations needed to generate each pseudorandom number. RSA-based cryptographic applications often use small exponents such as $e=3,5,17$.\cite{FergusonSchneierKohno2010}  Messages with $m^e < n$ and $(n-m)^e < n$ result in trivially decodable ciphertexts, so exponents $e < \log_2 n$ result in additional cribs. Cribs can be avoided by padding messages to ensure no messages are close to $0$ or $n$. For cryptographic applications, the padding needs to have a random component.\cite{Schneier1994, FergusonSchneierKohno2010} 
For our application, eliminating cribs from the message stream biases the ciphertext distribution due to the elimination of ciphertexts formed from messages with $m^e < n$ and $(n-m)^e < n$,
resulting in a slight but systematic under-sampling of small and large values of $c$. Even though the effect of this is small, we choose not to implement an algorithm that does not give a uniform distribution of ciphertexts over the full period. 
For our purpose, it is not necessary to eliminate cribs, since they would appear in a random message sequence, but rather to prevent correlated sequences of cribs.
Our goal is to select a simple skip pattern that ensures a uniform sampling of the set of all messages, is computationally fast, has a long period, allows the use of small encryption exponents, and avoids correlated cribs. 
We will accomplish this by using a pseudorandom skip sequence that, over the full period of the generator, uniformly samples the messages in $\mathbb{Z}/n\mathbb{Z}$, but we will first examine the properties of pseudorandom ciphertext sequences derived from simpler skip patterns.

\begin{itemize}

\item The simplest skip sequence that uniformly samples $\mathbb{Z}/n\mathbb{Z}$ is the unit skip, i.e. $s_k = 1$ for all $k$. Pohlig--Hellman encryption of this sequence is, in effect, a block cipher operating in counter mode.\cite{Schneier1994,FergusonSchneierKohno2010} 
The message sequence is $m_k=(m_0 + k) \thinspace\textrm{mod}\thinspace n$, with the period $P$ of the generator being $P=n$. For 32-bit moduli, one can easily test the full period of the generator.  In spite of the cribs near $m=0$ and $m=n-1$, the sequence $c_k = k^e \thinspace\textrm{mod}\thinspace n$ with $0\leqslant k <n $ passes most statistical tests for randomness even for small exponents. Naturally, the full-period sequence produces a perfectly uniform one-dimensional histogram since after $n$ steps every value $c$ will appear once, and only once, in the sequence. Except for the one-dimensional frequency test, and other tests that are affected by the uniformity of the sampling of numbers in the range $\mathbb{Z}/n\mathbb{Z}$, and in spite of the symmetry noted below, 
the exponentiation cipher with unit skips passes all of our other statistical correlations tests for exponents $e=9$ and $e=17$. By comparison, all linear congruential generators fail \textit{all} $D$-dimensional correlations tests once a substantial fraction of the period has been exhausted, since all $D$-dimensional histograms become uniform as the period of the generator is approached.\cite{knuth} The exponentiation cipher does not suffer from this; 
see figures 1 and 2 which display the two-dimensional correlation patterns for a prime number linear congruential generator, and a unit-skip Pohlig--Hellman generator using the same prime modulus. 
As with most pseudorandom number generators, there is a symmetry in the pseudorandom sequence. The Pohlig--Hellman cipher has the symmetry 
\begin{align*}
(n-m)^e \thinspace\textrm{mod}\thinspace n=n - m^e \thinspace\textrm{mod}\thinspace n ,
\end{align*} 
i.e. the ciphertexts derived from messages in $[1\thinspace .\thinspace .\thinspace (n-1)/2]$ are strongly correlated with the messages in $[(n+1)/2\thinspace .\thinspace .\thinspace (n-1)]$, but in reverse order.

\item The next simplest skip algorithm is the constant skip: $s_k = b$ with $1<b<n-1$, so $m_k=(m_0 + k b) \thinspace\textrm{mod}\thinspace n$. This message pattern is especially important for understanding the quality of our proposed pseudorandom skip algorithm discussed next. 
For most values of $b$, the constant skip removes sequential and closely spaced cribs. 
However, other than breaking up closely spaced cribs, the pattern produced by constant skips is not  substantially better than 
the unit skip pattern since 
\begin{align}\label{constantskipsymmetry}
(kb)^e\thinspace\textrm{mod}\thinspace n=\left( (b^e \thinspace\textrm{mod}\thinspace n)(k^e \thinspace\textrm{mod}\thinspace n)\right) \thinspace\textrm{mod}\thinspace n,
\end{align}
i.e. this is the same as the unit-skip sequence except for a single constant factor $b^e \thinspace\textrm{mod}\thinspace n$. 
Like the unit skip case, there is a symmetry in the pseudorandom sequence since $((n-k)b)^e \thinspace\textrm{mod}\thinspace n = n - (kb)^e\thinspace\textrm{mod}\thinspace n$. As with the unit skip, the constant skip passes a battery of  full-period statistical tests for $e=9$ and $e=17$, except ones affected by the uniformity of the one-dimensional distribution.

\item Our recommended skip pattern is a pseudorandom skip produced by a prime number linear congruential pseudorandom number generator:\cite{knuth, lecuyer1999, lecuyer1988}
\begin{align}\label{congruentialgenerator1}
s_k= a s_{k-1} \thinspace\textrm{mod}\thinspace p= s_0 a^k \thinspace\textrm{mod}\thinspace p ,
\end{align}
with prime modulus $p<n$. The  multiplier $a$ is chosen to be a primitive root\cite{Koshy2002, Silverman2006, Koblitz1987} $\textrm{mod}\thinspace p$ that  
delivers a full period, well-tested
pseudorandom sequence.\cite{knuth, lecuyer1999,lecuyer1988} 
This gives pseudorandom skips $s_k$ in the range $(\mathbb{Z}/p\mathbb{Z})^*$, with the period of the skip sequence being $p-1$.
This results in the following pseudorandom ciphertext sequence:
\begin{subequations}\label{pseudoskip}
\begin{align}
s_k &= s_0 a^k \thinspace\textrm{mod}\thinspace p , \\
m_k &= \left( m_0 + \sum_{j=1}^k{ s_0 a^j \thinspace\textrm{mod}\thinspace p } \right)\thinspace\textrm{mod}\thinspace n ,
\label{pseudoskipb}  \\
c_k &= m_k^e \thinspace\textrm{mod}\thinspace n .
\end{align}
\end{subequations}
Using a pseudorandom skip sequence serves several important purposes: 

\begin{itemize}

\item A pseudorandom skip effectively eliminates the problem of correlated cribs, allowing the use of small exponents. 

\item Using a pseudorandom skip extends the period of the generator to $P=n(p-1)$. 

\item The method provides a uniform sampling of ciphertexts over the full period of the generator. Each message $m$ in $\mathbb{Z}/n\mathbb{Z}$, and hence each ciphertext $c$ in $\mathbb{Z}/n\mathbb{Z}$, will appear exactly $p-1$ times in the sequence.

\item The state of each generator is defined by only seven integers: $\{m,s,c,n,e,p,a\}$.

\item Implementations using 32-bit primes are fast. With small exponents, each pseudorandom number requires only a few 64-bit multiply--mod operations that can be implemented in hardware on 64-bit processors.

\item The implementations suggested below pass a battery of statistical tests with up to $10^{13}$ pseudorandom numbers per test.

\end{itemize}

First, let's prove that the period of the generator is $P=n(p-1)$. Since $a$ is  
a primitive root mod $p$, after $p-1$ pseudorandom skips $s_k$ will have cycled through every value in $(\mathbb{Z}/p\mathbb{Z})^*$, so $s_{k+p-1}=s_k$ and $m_{k+p-1}=\left( m_k + \sum_{s=1}^{p-1}{s} \right) \thinspace\textrm{mod}\thinspace n = ( m_k + p(p-1)/2 ) \thinspace\textrm{mod}\thinspace n$.  This means the sequence of messages separated by multiples of $p-1$ steps in the sequence are derived from a constant skip $b=p(p-1)/2 \thinspace\textrm{mod}\thinspace n$. Since $p$ and $(p-1)/2$ are coprime to $n$, the constant skip $b$ is also co-prime to $n$. 
Therefore, the state of the generator after $k=q(p-1) + r$ steps, with $q=\lfloor k/(p-1)\rfloor$ and $r=k\thinspace\textrm{mod}\thinspace (p-1)$, is given by
\begin{subequations}\label{pseudogenerator1}
\begin{align}
s_k &= s_0 a^r \thinspace\textrm{mod}\thinspace p , \label{pseudogenerator1a}\\
m_k &= \left( m_0 + q p (p-1)/2  + \sum_{j=1}^r{ \left( s_0 a^j \thinspace\textrm{mod}\thinspace p\right) } \right)\thinspace\textrm{mod}\thinspace n  , \label{pseudogenerator1b}\\
c_k &= m_k^e \thinspace\textrm{mod}\thinspace n  \label{pseudogenerator1c}.
\end{align}
\end{subequations}
This demonstrates the mathematical form of the pseudorandom sequence across the full period of the generator. Since $s_{k+n(p-1)}=s_k$ and $m_{k+n(p-1)} = m_k$, and each subsequence of messages of length $p-1$ is different from every other subsequence, the period the generator is $P=n(p-1)$. Over the full period, every ciphertext in $\mathbb{Z}/n\mathbb{Z}$ will appear exactly $p-1$ times. 
One can use equation \eqref{pseudogenerator1} to skip $k$ steps forward directly to any point in the pseudorandom sequence, with skips being fast if $k$ is close to a multiple of $p-1$. Backward skips of are accomplished by replacing $a$ with $a^{-1} \thinspace\textrm{mod}\thinspace p$, subtracting rather than adding in equation \eqref{pseudogenerator1b}, and reordering  equations \eqref{pseudogenerator1a} and \eqref{pseudogenerator1b}.\cite{plessthann}

The simple form of the ciphertext sequence allows one to determine the full-period $D$-dimensional correlations pattern for rectangular regions with volume $L_0 L_1^{D-1}$ in $O(p L_0)$ steps, i.e. without needing to exhaust the generator. The $D$-dimensional density of points $(c_k,c_{k+1},\ldots , c_{k+D-1})$ over the full period is $\rho_D = (p-1)/n^{D-1}$ so the average number of points in the volume above is $(p-1)L_0 (L_1/n)^{D-1}$. One can select $L_0$ points $c_k$, with the succeeding points given by $c_{k+1}=(c_k^d \thinspace\textrm{mod}\thinspace n + s)^e \thinspace\textrm{mod}\thinspace n$, $c_{k+2}=(c_k^d \thinspace\textrm{mod}\thinspace n + s + (a s)\thinspace\textrm{mod}\thinspace p)^e \thinspace\textrm{mod}\thinspace n$, etc.~where the skips $s$ take on all values in $(\mathbb{Z}/p\mathbb{Z})^*$.  A magnified region near the origin of the full-period two-dimensional pattern for safe prime $n=4294967087$, $p=2147483647$, $a=784588716$,\cite{lecuyer1999} and $e=9$ is shown in figure 3, with the full-period three-dimensional pattern shown in figure 4.

We tested the quality of 32-bit pseudorandom sequences based on pseudorandom skip sequences using a battery of independent statistical tests.  We first tested uniform double precision floating point pseudorandom sequences $R_k$ over the period of the skip generator for more than ten-thousand different safe primes between $2^{31}$ and $2^{32}$, and used a single prime number linear congruential pseudorandom number generator recommended by L'Ecuyer:\cite{lecuyer1999} $p=2^{31}-1=2147483647$ and $a=784588716$. Even using exponents as small as $e=3$, the intrastream pseudorandom sequences pass all of our statistical tests across the period of the skip generator. 

We then used equation \eqref{pseudogenerator1} to test for intrastream correlations among ciphertexts separated by $p-1$ steps in the sequence across the constant skip period $P=n$ using thousands of safe primes between $2^{31}$ and $2^{32}$. This pseudorandom sequence is  given by $c_{q(p-1)}= (m_0+ q b)^e \thinspace\textrm{mod}\thinspace n$, with $b=p (p-1)/2 \thinspace\textrm{mod}\thinspace n$ and $q=0,1,\thinspace . \thinspace . n-1$. These constant skip sequences pass all of our the statistical tests for exponents as small as $e=7$ for up to $2^{25}$ pseudorandom numbers, and all but the one-dimensional tests across the period $P=n$. (As noted earlier, the one-dimensional constant skip distribution becomes uniform over the period. These are the only tests that the constant skip sequences appear to fail.) 
Based on this, we recommend using exponents $e=9$ or $e=17$, which require only four or five multiply--mod operations per exponentiation, respectively.  

Since ciphertexts separated by $k<p-1$ steps demonstrate good intrastream statistics because of the pseudorandom skip, and ciphertexts separated by multiples of $p-1$ steps demonstrate good statistics with constant skips $b=p(p-1)/2 \thinspace\textrm{mod}\thinspace n$, one has good reason to believe that the entire pseudorandom sequence should pass a battery of statistical tests until the length of the pseudorandom sequence approaches the full period. To test this, we performed intrastream statistical tests of the pseudorandom 32-bit skip algorithm over as large a fraction of the period and for as many different moduli as possible. We tested sequences of $10^{11}$ pseudorandom numbers using hundreds of different primes, sequences of $10^{12}$ pseudorandom numbers using dozens of different primes, and sequences of $10^{13}$ pseudorandom numbers using a few different primes. This latter test corresponds to several thousand periods of the skip sequence. The method passed every test we applied.

We also tested to ensure that the algorithm displayed lack of correlation between streams. Suppose one has $N_p$ processes each with a different prime modulus $n^{(\alpha)}$ with $\alpha=0,1,\ldots ,N_p-1$, and pseudorandom sequences $R_0^{(\alpha)}, R_1^{(\alpha)}, \ldots$. Our interstream correlations tests draw the pseudorandom numbers  from the $N_p$ streams in the order $R_1^{(0)}, R_1^{(1)}, R_1^{(2)}, \ldots , R_1^{(N_p-1)}, R_2^{(0)}, R_2^{(1)}, \ldots $. We tested  groupings of $N_p=2$, $32$, $1024$, $32768$ and $1048576$ different streams, using exponents $e= 5, \medspace 9, \medspace \mbox{and} \medspace 17$.  We also included tests using $N_p=3,060,794$ safe primes, i.e. all of the safe primes in  $[2^{31} \thinspace .\thinspace . \thinspace 2^{32}]$. To test that seeding coincidences do not create spurious correlations, we performed many of the interstream tests by starting every sequence with exactly the same initial values $m_0$ and $s_0$, but avoiding cases in which one of the early messages in the sequence is a crib. The interstream correlations passed every test for up to $10^{13}$ pseudorandom numbers per test. 

\end{itemize}

\section{Tests}
We first applied well-established pseudorandom number test suites DIEHARD,\cite{diehard} NIST\cite{NIST}, and TestU01,\cite{U01Test} to ensure the generator passes a wide variety of tests. 
We then applied the following ten additional tests that produce histograms to which one can apply a chi-square test.\cite{knuth} For each test, we calculated $\chi^2$ and the associated $\mathscr{P}$-value, i.e. the one-sided probability of $\chi^2$ having that value above or below the median. We applied these additional tests to sequences of up to $10^{13}$ pseudorandom numbers per test. 

\begin{itemize}

\item One-dimensional frequency test:\cite{knuth} We distributed the sequence of pseudorandom numbers into a one-dimensional histogram with $2^{20}$ bins, and compared the histogram to a uniform Poisson distribution. 

\item Two-dimensional serial test:\cite{knuth} We distributed pairs of pseudorandom numbers into a two-dimensional histogram with $2^{20}$ bins, and compared the histogram to a uniform Poisson distribution. This tests for sequential pair correlations in the sequence.

\item Three-dimensional serial test:\cite{knuth} We distributed triplets of pseudorandom numbers into a three-dimensional histogram with $10^6$ bins, and compared the histogram to a uniform Poisson distribution. This tests for sequential triplet correlations in the sequence.

\item Four-dimensional serial test:\cite{knuth} We distributed groups of four pseudorandom numbers into a four-dimensional histogram with $2^{20}$ bins,  and compared the histogram to a uniform Poisson distribution. This tests for sequential four-point correlations in the sequence.

\item Five dimensional serial test:\cite{knuth} We distributed groups of five pseudorandom numbers into a five-dimensional histogram with $2^{20}$ bins,  and compared the histogram to a uniform Poisson distribution. This tests for sequential five-point correlations in the sequence.

\item Poker test:\cite{knuth} We used groups of five pseudorandom numbers and counted the number of pairs, three-of-a-kind etc.~formed from five cards and ten denominations, and compared the resulting histogram to a Poisson distribution derived from the exact probabilities. This tests for a variety of five-point correlations in the sequence.

\item Collision test:\cite{knuth} We distributed $2^{14}$ pseudorandom numbers into $2^{20}$ bins and compared the histogram of the number of collisions against a Poisson distribution derived from the exact probabilities. This tests for long-range correlations in the sequence.

\item Runs test:\cite{knuth} We compared the histogram of the length of runs of 0's ($R\leq 0.5$) and 1's ($R>0.5$) to the Poisson distribution derived from the exact probabilities. This tests for short-range correlations of the leading bits.

\item Fourier test:\cite{MascagniSrinivasan2000} We used a fast Fourier transform\cite{fft} to calculate the Fourier coefficients of sequences of  $M=2^{20}$ double precision floating point pseudorandom numbers, 
\begin{align}
\hat{x}_k = \frac{1}{\sqrt{M}} \sum_{j=0}^{M-1} { x_j e^{2\pi i j k/M} } ,
\end{align}
where $x_j=R_j - 0.5$.
The real and imaginary parts of the Fourier coefficients $\hat{x}_k$ with $k=0,\ldots, M/2 - 1$ should each be gaussian distributed about zero with variance $1/24$.  We distributed the real and imaginary parts of the Fourier coefficients into a histogram, and compared the result to a Poisson distribution derived from the exact gaussian distribution. This tests for long-range pair correlations in the sequence. 

\item Two-dimensional Ising model energy distribution test:\cite{beale1996, pathriabeale2011} We performed a Wolff algorithm\cite{wolff} Monte Carlo simulation at the critical point of the two-dimensional Ising model on a $128\times128$ square lattice, and compared the energy histogram to a Poisson distribution derived from the exact probabilities\cite{beale1996, pathriabeale2011} Since the Wolff algorithm is based on stochastically growing fractal critical clusters that can span the system, this tests for long-range correlations in the pseudorandom sequence, and has proven to be effective at identifying weak generators.\cite{beale1996, ferrenberglandauwong1992} Assigning an independent generator to each of the 32768 bonds in the lattice tests provides an additional test for exposing interstream correlations.

\end{itemize}

For several of these tests, we sampled the low-order bits to confirm they do not harbor any hidden correlations. We define passing our tests by there being no $\mathscr{P}<10^{-8}$ events among the tests. Since there were tens of thousands of independent tests, we also counted the number of $\mathscr{P}<10^{-4}$ events in our samples to confirm that the number was consistent with the expected number. Finally, we applied chi-square tests to histograms of the full-period distribution of $(c_k, c_{k+1}, .. , c_{k+D-1})$ sequences in $2^{24}$  $D$-dimensional rectangular volumes of size $L_0 L_1^{D-1}$ with $L_0=2^7$ and $L_1\approx n/2^{24/(D-1)}$, for $D=3,4,5$. The generator passed every test.

\section{Implementation}

The algorithm for generating uniformly distributed double precision floating point pseudorandom numbers $R$ on the open interval $(0\thinspace .\thinspace .\thinspace 1)$ is:
\begin{subequations}\label{algorithm1}
\begin{align}
&s := (a s) \thinspace\textrm{mod}\thinspace p \\
&m := (m + s )\thinspace\textrm{mod}\thinspace n \\
&c := m^e \thinspace\textrm{mod}\thinspace n \label{CodeExponentiationStep}\\
&R := (c+1) / (n+1) \label{CodeFloatingPointStep}\\
&\textrm{RETURN}\medspace R 
\end{align}
\end{subequations}

In a 32-bit implementation, operations (9a)-(9c) of the the algorithm can be implemented using 64-bit unsigned integer arithmetic, which can be executed in hardware on 64-bit processors. 
This algorithm generates the sequence given by equations \eqref{pseudogenerator1}. For efficiency, one can precalculate the double precision floating point value $1.0/(n+1)$ and perform a floating point multiplication instead of a floating point divide in step \eqref{CodeFloatingPointStep}.

For a multiprocessor environment,  
each of the $N_p$ processes 
can be assigned an independent prime modulus $n^{(\alpha)}$
using well-established primality tests. The Rabin-Miller test,\cite{Koblitz1987, millerprime, rabinprime} which is the same as Algorithm P in Knuth,\cite{knuth} provides a simple probabilistic test for primality. Every odd prime $n=1+2^r t$ with $t$ odd satisfies one of the following conditions for every base $g$ in $(\mathbb{Z}/n\mathbb{Z})^*$: either $g^t \thinspace\textrm{mod}\thinspace n = 1 $, or $g^{2^j t} \thinspace\textrm{mod}\thinspace n = n-1$ for some some $j$ in the range $0 \leq j < r$. A composite number $n$ satisfying these criteria is called a strong pseudoprime to base $g$.  
For any odd composite, the number of bases $g$ 
for which $n$ is a strong pseudoprime to the base $g$ 
is less than $ n/4 $, so if the test is applied repeatedly with $M$ randomly chosen bases in $(\mathbb{Z}/n\mathbb{Z})^*$, the probability that a composite will pass every test is less than $4^{-M}$.\cite{Koblitz1987, knuth, millerprime, rabinprime} Better yet, the Rabin-Miller test can quickly and deterministically identify all primes below $2^{64}$. 
There are no composite numbers below $2^{64}$ that are strong pseudoprimes to all of the 
twelve smallest prime bases ($g=2,3,5,7,11,13,17,19,23,29,31,37$).\cite{PomeranceSelfridgeFlagstaff1980, http://oeis.org/A014233, Zhang2001, JiangDeng2012, wolframrabinmiller}
Therefore, any number less than $2^{64}$ that passes the Rabin-Miller test for all twelve of these bases is prime. 
Likewise, any number less than $2^{32}$ that passes the Rabin-Miller test for all of the five smallest prime bases ($g = 2, 3, 5, 7, 11$) is prime. For efficiency, one first checks to see if any small primes divide $n$ before applying the Rabin-Miller test.

One can initialize the generators using the system time variable 
and the process identifier $\alpha$ to construct unique values of $n^{(\alpha)}$, and different starting values 
$s_0^{(\alpha)}$ and $m_0^{(\alpha)}$ for each of the $N_p$ processes. One might use equation \eqref{pseudogenerator1} to skip each generator forward relative to some base state such as $m_0=0$ and $s_0=1$ to ensure the initial skips and messages are all unsynchronized. Even if two generators later become synchronized with the same values of $m$ and $s$, the values for the ciphertexts for different moduli will be different due to the encryption step, and the message synchrony is removed after a few skips, even if the skips were to remain synchronized. 

The implementation for 32-bit moduli is simple and fast since the multiply--mod operations can be executed in hardware using standard 64-bit unsigned arithmetic on 64-bit processors. The quality of the pseudorandom sequences does not appear to be dependent on the values of the prime moduli, but we recommend using safe primes selected from $[2^{31} \thinspace . \thinspace . \thinspace 2^{32}]$ unless the number of processes exceeds 3,060,794, the number of safe primes in that range.  If the implementation requires more than 3,060,794 instances, there are 49,091,941 primes $n$ in $[2^{31}\thinspace . \thinspace . \thinspace 2^{32}]$ with $e=9$ coprime to $n-1$, and 92,045,560 primes in that range with $e=17$ coprime to $n-1$. 
The number of well-tested 31-bit prime number pseudorandom skip generators is not scalable, but one can use the ones available in the literature\cite{lecuyer1988,lecuyer1999} to substantially extend the number of possible instances.\cite{poweroftwoskips}

The full period of each generator is $P=n(p-1) > 2^{62} \simeq 4.6 \times 10^{18}$. This 32-bit implementation passes all of our intrastream and interstream correlations tests for $e=9$ and $e=17$, for up to $10^{13}$ pseudorandom numbers per prime modulus. The exponent $e=9$ requires only five multiply--mod operations per pseudorandom number, one for the skip and four for the exponentiation, and the exponent $e=17$ requires only six. Consequently, the code is compact and simple enough to be implemented as an in-line function.

If far longer periods or far more instances are needed, one can implement the algorithm using 
64-bit primes. There are approximately $3\times 10^{15}$ safe primes in $[2^{63} \thinspace . \thinspace . \thinspace 2^{64}]$. L'Ecuyer\cite{lecuyer1999} provides suitable pseudorandom skip parameters near $2^{63}$  so the period of each process is $P=n(p-1)>2^{126}\simeq 8.5 \times 10^{37}$. Using 64-bit prime moduli results in a speed penalty with current processors since the multiply--mod operations need to be implemented in using 128-bit unsigned integers. 
Since a single process is unlikely to exhaust a 63-bit pseudorandom skip, one might consider using smaller encryption exponents for efficiency. Preliminary statistical tests indicate the method works well for 64-bit safe prime moduli even when using $e=3$, the smallest allowed exponent. 

\section{Conclusion}

We propose a new  class of parallel pseudorandom number generators based on Pohlig--Hellman exponentiation ciphers. The method creates pseudorandom streams by encrypting simple sequences of integer messages. The method is fully scalable based on parametrization since each process can be assigned a unique prime modulus. By using pseudorandom skips among messages, one can use small exponents and the period is greatly extended. For 32-bit implementations, only a few 64-bit multiply--mod operations are needed per pseudorandom number. There are millions of possible independent instances, and the period of each instance is is greater than $10^{18}$. We have tested thousands of different pseudorandom streams for intrastream and interstream correlations using up to $10^{13}$ pseudorandom numbers per test, and all pass a battery of statistical tests. A 64-bit implementation would have more than $10^{15}$ possible instances and periods greater than $10^{37}$. Sample C++ code for a 32-bit multi-processor MPI implementation of the Pohlig--Hellman pseudorandom number generator with pseudorandom skip can be found at \url{http://works.bepress.com/paul_beale/}.

\section{Acknowledgements} We thank Matt Glaser, Rudy Horne, Nick Mousouris, Ethan Neil, Robert Blackwell, John Black, and David Grant for helpful discussions. This work utilized the Janus supercomputer, which is supported by the National Science Foundation (award number CNS-0821794) and the University of Colorado Boulder. The Janus supercomputer is a joint effort of the University of Colorado Boulder, the University of Colorado Denver, and the National Center for Atmospheric Research.

\begin{figure}[h!]
\includegraphics[width=5in]{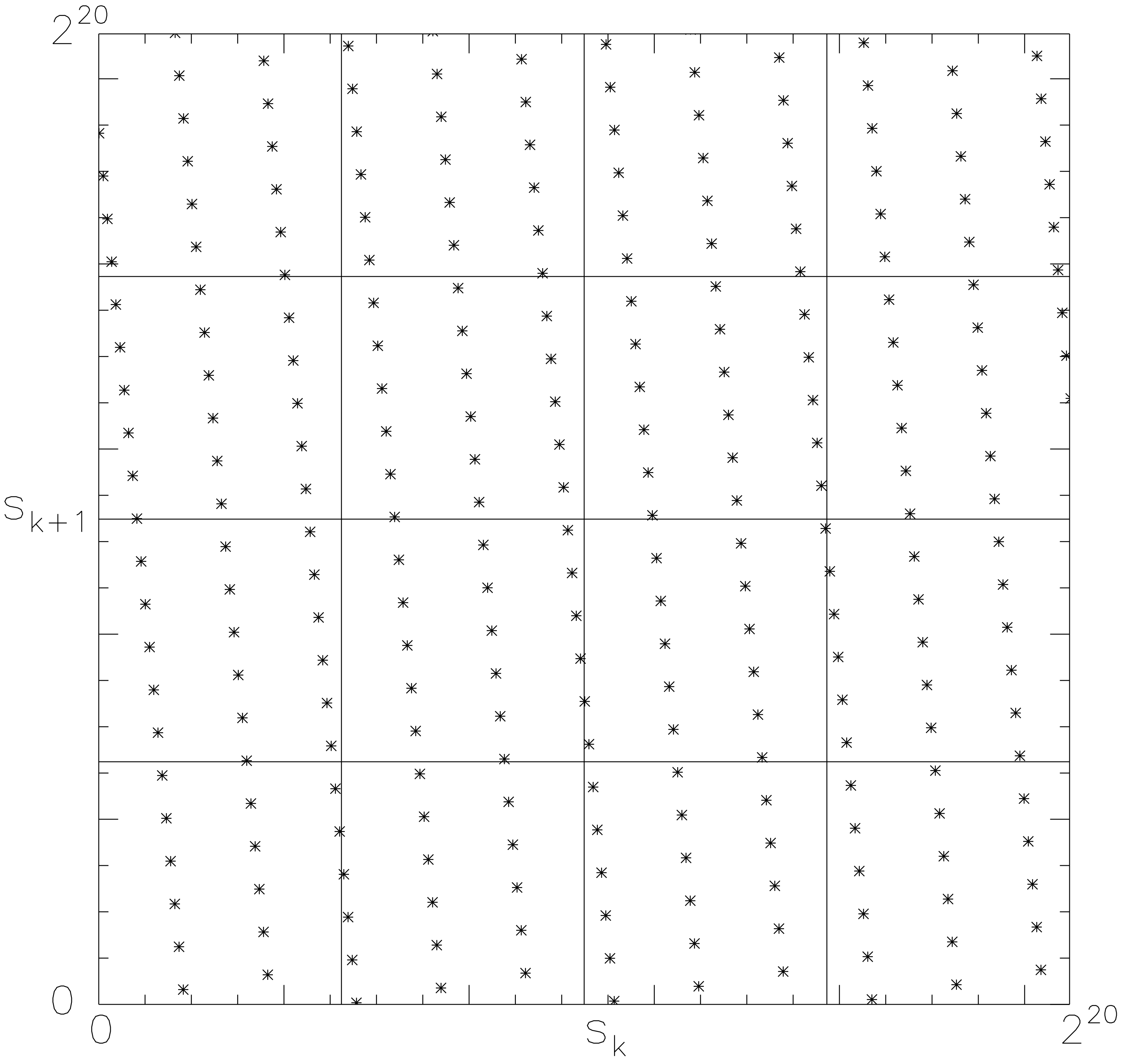}
\caption{Magnified region near the origin of the two-dimensional full-period $s_{k+1}$ vs. $s_k$ pattern delivered by a prime number linear congruential pseudorandom number generator (equation \eqref{congruentialgenerator1}) with $p=2^{32}-5=4294967291$ and $a=279470273$.\cite{lecuyer1999} The axes cover the range $[0\thinspace . \thinspace . \thinspace\thinspace 2^{20}]$, i.e. a linear magnification factor of 4096. The 16 subsquares have area $\Delta x^2=(2^{18})^2$ each. As with all linear congruential generators, every $D$-dimensional full-period pattern forms a perfect lattice,\cite{knuth} so the number of points in each cell is very close to $N_{cell}=\Delta x^2 / (p-1) \simeq  16$ events in this case.}
\end{figure}

\begin{figure}[h!]
\includegraphics[width=5in]{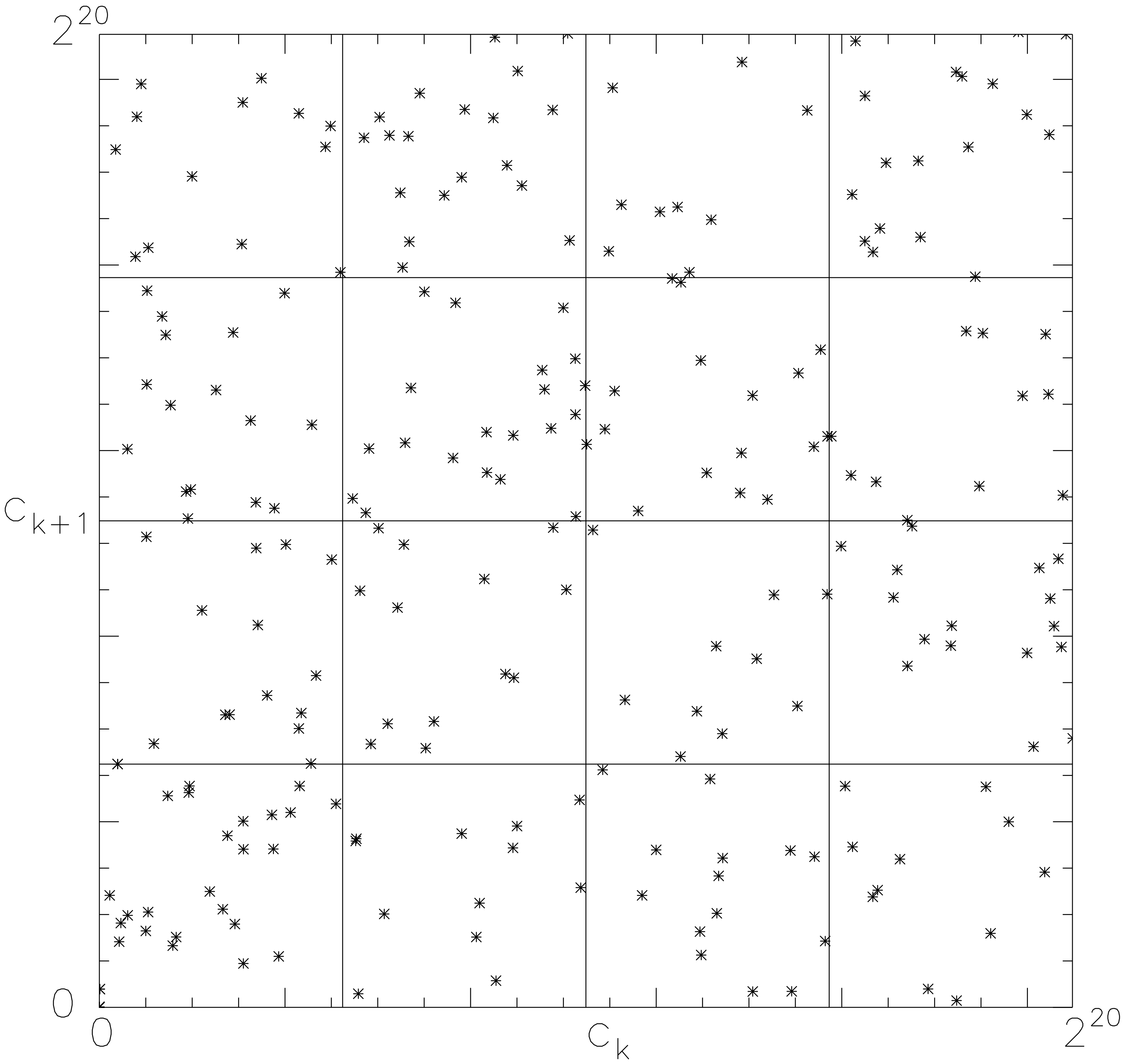}
\caption{Magnified region near the origin of the two-dimensional full-period $c_{k+1}$ vs. $c_k$ pattern for a Pohlig--Hellman exponentiation cipher (equation \eqref{basealgorithm}) using the same prime as in Fig. 1,  $n=2^{32}-5=4294967291$, exponent $e=9$, and unit skip, i.e. $m_k=0, 1,2,\ldots , n-1$. The axes cover the range $[0\thinspace . \thinspace . \thinspace\thinspace 2^{20}]$, i.e. a linear magnification factor of 4096. The 16 subsquares have area $\Delta x^2=(2^{18})^2$ each. Unlike linear congruential generators, the distribution of points in the cells approximate a Poisson distribution with $N_{cell} \approx  \Delta x^2 /n  \pm \sqrt{\Delta x^2 / n }\simeq 16\pm 4$ in this case. The unit-skip correlated cribs $(0,1)$, \ $(1,2^9=512)$,  $(512,3^9=19683)$, and $(19693,4^9=2^{18}=262144)$ appear near the vertical axis in the lower left square.}
\end{figure}

\begin{figure}[h!]
\includegraphics[width=4.5in]{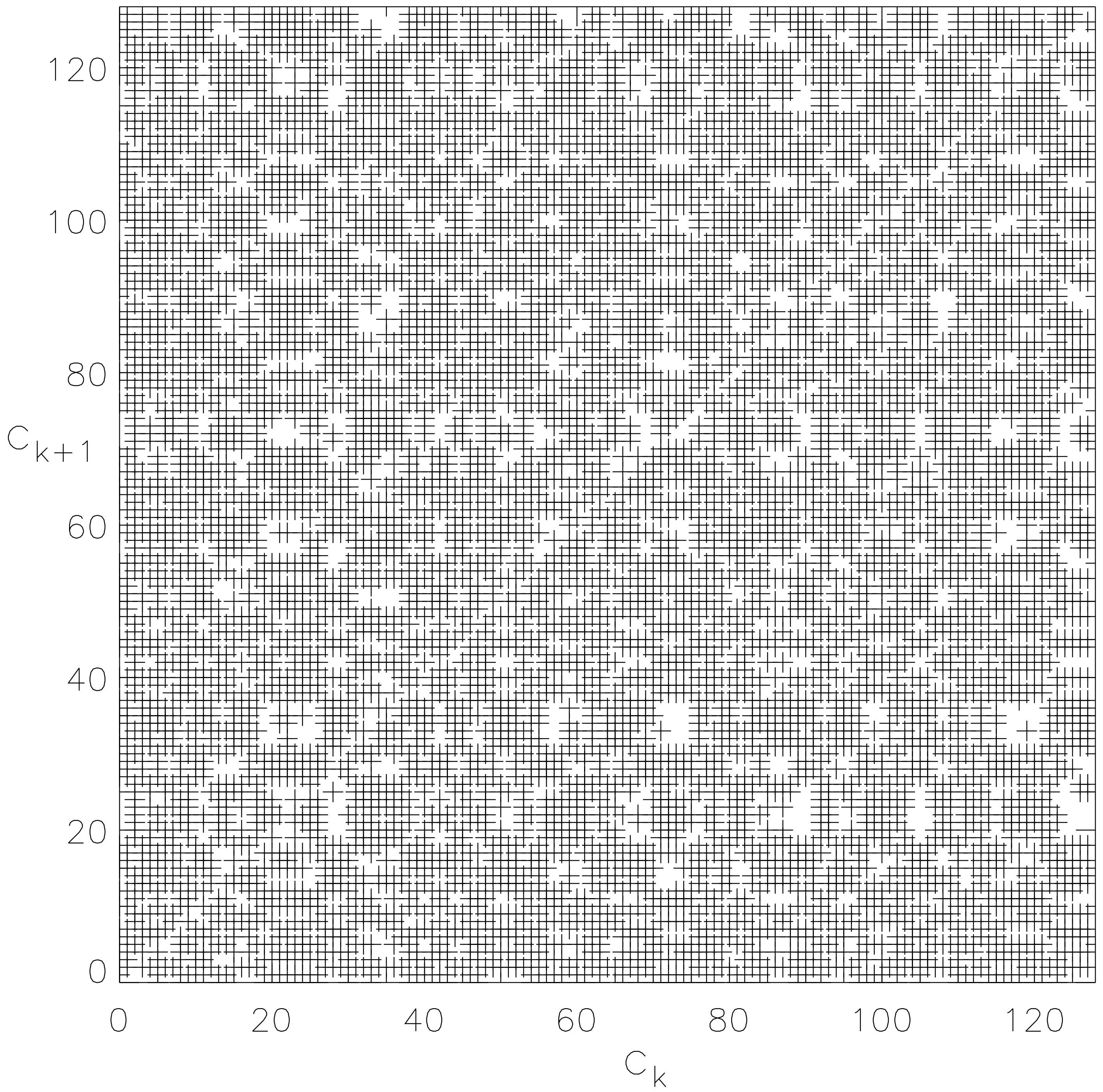}
\caption{
Sequential ciphertext pairs $(c_k,c_{k+1})$ in a magnified two-dimensional square region over the full period of the generator for a pseudorandom skip with $n=4294967087$, $p=2147483647$,   
$e=9$ and $d=3817748521$. The region shown is a $2^{7} \times 2^{7}$ square closest to the origin,
Every prime $n$--$e$--$p$ combination gives a different full-period pattern, but the two-dimensional full-period pattern is independent of the value of primitive root $a$. The average full-period density of the pairs in two dimensions is $\rho_2=(p-1)/n\simeq 0.5$. Each integer pair possibility appears either once or not at all. 
The absence of any points along the diagonal with $c_{k+1}=c_k$ is because $0<s_k<p$ for all $k$. Even though the full period is $P=n(p-1)\simeq 4.6\times 10^{18}$, it is feasible to determine the local  full-period pattern without exhausting the generator since $c_{k+1}=((c^d_{k}\thinspace\textrm{mod}\thinspace n) +s_{k+1})^e \thinspace\textrm{mod}\thinspace n$. One selects only ciphertexts $c_{k}$ in the chosen domain while testing all  skips $s_{k+1}$ in 
$(\mathbb{Z}/p\mathbb{Z})^*$
to see which ones give $c_{k+1}$'s in the chosen range.  Primes with $p \approx n$ give patterns that fill in nearly every $(c_k,c_{k+1})$ pair, except those along the diagonal.
}
\end{figure}

\begin{figure}[h!]
\includegraphics[width=5in]{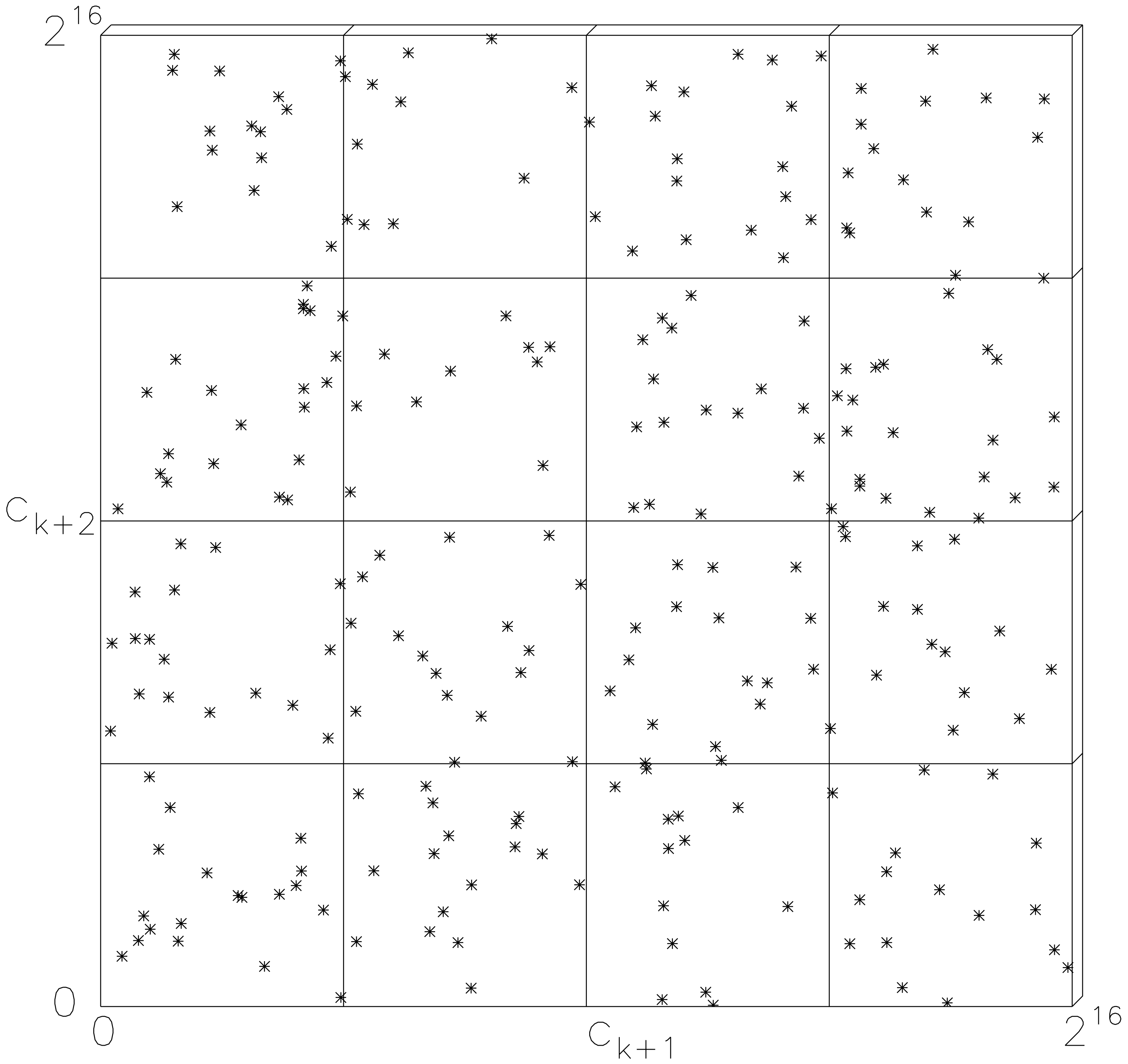}
\caption{Sequential ciphertext triplets $(c_k,c_{k+1},c_{k+2})$ over the full period of the generator in a rectangular three-dimensional region for a pseudorandom skip with $n=4294967087$, $p=2147483647$,  $a=784588716$, 
$e=9$ and $d=3817748521$.  The region shown is a $L_0 L_1^{2}=2^9 \times 2^{16} \times 2^{16}$ rectangular slab  closest to the origin with volume $V_3=2^{41}$. The short first dimension is projected onto the front face. 
Even though the full period is $P=n(p-1)\simeq 4.6\times 10^{18}$, it is feasible to determine the local  full-period pattern without exhausting the generator since $c_{k+1}=((c^d_{k}\thinspace\textrm{mod}\thinspace n) +s_{k+1})^e \thinspace\textrm{mod}\thinspace n$ and 
$c_{k+2}=((c^d_{k}\thinspace\textrm{mod}\thinspace n) +s_{k+1} + (a s_{k+1}) \thinspace\textrm{mod}\thinspace p)^e \thinspace\textrm{mod}\thinspace n$. One selects all ciphertexts $c_{k}$ in $[0 \thinspace .\thinspace . \thinspace L_0-1]$ and all  skips $s_{k+1}$ in $(\mathbb{Z}/p\mathbb{Z})^*$ to see which ones give $c_{k+1}$ and $c_{k+2}$ inside the chosen volume. The three-dimensional density of triplets is $\rho_3=(p-1)/n^2$. The result shown here of the occupancies of the sixteen subslabs is consistent with a Poisson distribution with  $N_{cell} \approx \rho_3 V_{cell} \pm \sqrt{\rho_3 V_{cell}} \simeq 16\pm 4$.  
}
\end{figure}


\begin{thebibliography}{5}
\bibitem{poweroftwocongruentialweaknesses} The period of every such generator is $p$, and the low order bits are highly correlated within each stream, as well as between streams.
\bibitem{plessthann}The mathematical form of the message sequence can be used to demonstrate why the method requires $p<n$. If $n=p$, one can evaluate the sum in equation \eqref{pseudoskipb} in closed form to give 
$m_k = (m_0 + s_0 a(a-1)^{-1}(a^k-1))\thinspace\textrm{mod}\thinspace p$. 
Therefore, using Fermat's little theorem, $m_{k+p-1}=m_k$ and $s_{k+p-1}=s_k$, so period is reduced to $P=p-1$.
\end{thebibliography}
\end{document}